\documentclass[prd,twocolumn,showpacs,showkeys,preprintnumbers,floatfix,
nofootinbib,superscriptaddress]{revtex4-1}
\usepackage{amsfonts} 
\usepackage{amssymb} 
\usepackage{amsmath} 
\usepackage{graphicx} 
\usepackage{subfigure} 
\usepackage{array} 
\usepackage{dcolumn} 
\usepackage{bm} 
\usepackage{latexsym} 
\usepackage{longtable} 
\usepackage{hyperref} 
\usepackage{verbatim}
\usepackage{epsfig}
\usepackage{color}
\newcommand{\cms}[1]{}

\newcommand{\CF}{{\cal F}}

\newcommand{\CM}{{\cal M}}
\newcommand{\CX}{{\cal X}} 
\newcommand{\CH}{{\cal H}} 
\newcommand{\CY}{{\cal Y}}

\DeclareMathOperator{\ttt}{t}
\DeclareMathOperator{\ccc}{c}
\DeclareMathOperator{\sss}{s}

\newcommand{\MeV}{\mathop{\rm MeV}\nolimits}

\renewcommand{\Re}{\mathrm{Re\,}}

\begin{document}


\title{Multiple-channel generalization of Lellouch-L\"uscher formula}
\author{Maxwell T. Hansen}
\email[Email: ]{mth28@uw.edu}
\affiliation{
 Physics Department, University of Washington, 
 Seattle, WA 98195-1560, USA \\
}

\author{Stephen R. Sharpe}
\email[Email: ]{srsharpe@uw.edu}
%
%
\affiliation{
 Physics Department, University of Washington, 
 Seattle, WA 98195-1560, USA \\
}
%
%
%
%
%
%
\date{\today}
\begin{abstract}
We generalize the Lellouch-L\"uscher formula, relating
weak matrix elements in finite and infinite volumes, to the case of
multiple strongly-coupled decay channels into two scalar particles.
This is a necessary first step on the way to a lattice QCD calculation of
weak decay rates for processes such as $D\to\pi\pi$ and $D\to K \overline K$.
We also present a field theoretic derivation of the generalization of
L\"uscher's finite volume quantization condition to multiple two-particle 
channels. We give fully explicit results for the case of two channels,
including a form of the generalized Lellouch-L\"uscher formula 
expressed in terms of derivatives of the energies of finite volume
states with respect to the box size.
Our results hold for arbitrary total momentum and for degenerate or
non-degenerate particles.
\end{abstract}
\maketitle

\section{Introduction\label{sec:intr}}

Lattice calculations have made considerable progress
toward a first-principles determination of the $K\to\pi\pi$ weak decay
amplitudes~\cite{Blum:2011,Blum:2011n2}.
The methodology is now in place, results for the
$I=2$ final state with a complete error budget 
are available~\cite{Blum:2011n2},
and complete results for the more challenging $I=0$ 
final states should become available in the next few years.
At that stage we will finally learn whether and in what manner QCD
can explain the $\Delta I=1/2$ rule and the observed CP-violation
rate in $K\to\pi\pi$ decays.

Encouraged by this progress, it is natural to consider what
information lattice calculations might eventually offer concerning the
decays of heavier mesons. For example, the LHCb experiment recently
reported evidence for CP-violation in (the difference of)
$D^0\to\pi^+\pi^-$ and
$D^0 \to K^+ K^-$ decays~\cite{LHCb:2011}.
Although the rate is larger than naive expectations from
the Standard Model (SM), there is, at present, sufficient uncertainty in the
SM prediction for it to be consistent with the LHCb result
(see, e.g. Refs.~\cite{Golden:1989,Isidori:2011,Brod:2011,Bhattacharya:2012,Franco:2012,Zupan:2012}).
This raises the obvious question of whether a calculation using lattice
methods is feasible. 

The aim of this paper is to take a first
step in developing the methodology for such a calculation.
We show how, if one can ignore all but two-particle channels, then
a generalization of the work of L\"uscher, and of Lellouch and L\"uscher,
would allow, in principle, a calculation of the required matrix elements
from lattice calculations in a finite volume.
In practice, however, channels with more than two particles
are coupled by the strong interactions to $\pi\pi$ and $K\overline K$,
e.g. the four pion channel, and they cannot be ignored at center of mass (CM) energies as high as the \(D^0\)-meson mass ($M_{\! D^0} = 1865 \MeV$).
Thus our method would yield only semi-quantitative results
for the desired matrix elements. Nevertheless, it is a necessary first step,
and work is underway to extend the methodology to channels with
multiple particles (see, e.g., Ref.~\cite{Polejaeva:2012}).

It is instructive to recall the
three essential ingredients needed for the lattice
calculation of $K\to\pi\pi$ amplitudes.
First, one needs to know the relation between the energies of
two-pion states in a finite box and the infinite-volume scattering
amplitude. This was worked out by L\"uscher in 
Refs.~\cite{Luescher:1986n1, Luescher:1986n2, Luescher:1991n1,Luescher:1991n2}
(and generalized to a moving frame in 
Refs.~\cite{Kari:1995,Kim:2005,Christ:2005}).
Second, one needs the relation between the matrix element that
one can determine on the lattice, which connects a kaon to a finite-volume
two pion state, and the infinite volume matrix element which determines the
decay rate. This was provided by Lellouch and L\"uscher in 
Ref.~\cite{Lellouch:2000} (and generalized to a moving frame
in Refs.~\cite{Kim:2005,Christ:2005}).
Finally, one must calculate the large number of Wick
contractions that contribute, including several quark-disconnected
contractions requiring special methods and high statistics. 
In this stage one also extrapolates to physical quark masses.
This entire program has been carried out for the $I=2$ final 
state~\cite{Blum:2011n2}, and a successful pilot calculation has been done
for the more challenging $I=0$ case~\cite{Blum:2011}.

The calculation of \(D^0\) decays is considerably more
challenging. In particular, the first two of the 
three aforementioned ingredients need to be generalized to account for the
opening of many channels.
If we focus on the $I=0$ final state,
then strong-interaction rescattering connects two-pion final states
to those with four, six, etc. pions, as well as
$K\overline K$ and $\eta\eta$ states.\footnote{%
It is important to note that the fact that the $D^0$ has a very
large number of decay channels~\cite{rpp:2010} is not itself
a concern, but rather that, having fixed the final-state quantum
numbers, in our case to $I=0$, there are still a large number
of states. In a lattice calculation, one can separately consider the
decays to states with differing strong-interaction quantum numbers.}
As already noted, we consider here only the case in which several
two-particle channels are open, which for the $D^0$ would mean keeping
the $\pi\pi$, $K\overline{K}$ and $\eta\eta$ channels while ignoring
those with four or more pions. We make this approximation not because
we think that it is a good description of reality at the $D^0$ mass,
but rather because it is a necessary first step towards the required formalism.

Within this approximation, 
we provide here the generalization
of both the L\"uscher quantization condition and
the Lellouch-L\"uscher (LL) formula.
These generalizations are useful also in many other systems.
For example,
the quantization condition allows the determination of
the parameters of the $S$-matrix in the $I=0$ channel above the
two kaon threshold (and thus in the region of the $f_0(980)$ resonance),
because the coupling to four or more pions remains weak for such energies.
The same should be true in the $I=1$ case, where $K\overline K$ and
$\eta\pi$ are the dominant channels in the vicinity of the $a_0(980)$.
The multi-channel LL formula can be used to calculate
\(K \rightarrow \pi \pi\) amplitudes including isospin breaking 
(so that $\pi^{\pm}$ and $\pi^0$ are not degenerate).
Generalization to baryon decays are also possible, but this requires
dealing with particles with spin, which we do not attempt here.

There have been a large number of recent papers studying the
generalization of the L\"uscher quantization condition to
multiple two-body channels~\cite{Liu:2005,Lage:2009,Bernard:2010,Doering:2011}
and assessing its utility.
The work of Ref.~\cite{Liu:2005} uses non-relativistic quantum mechanics,
while Ref.~\cite{Lage:2009} is based on a non-relativistic 
effective field theory. References~\cite{Bernard:2010} 
and \cite{Doering:2011} are based on relativistic field theory,
and give an explicit result 
[Eq.~(3.5) of Ref.~\cite{Bernard:2010}]
for the case of two s-wave channels
in which the total momentum vanishes and in which the
contributions from higher partial waves are assumed negligible.
We also note that the multiple channel problem has been studied
using an alternative approach based on 
the Bethe-Salpeter wavefunction~\cite{Aoki:2011}.

We provide here, as a step on the way to the generalized LL formula,
a derivation of the multiple-channel quantization condition within
quantum field theory.
We include all allowed mixing between different partial waves.
No assumptions about the form of
the interactions are needed,
aside from the proviso, common to all approaches,
that the range of the interaction must
be smaller than the box size.
Also, our result holds for any value of the total 
momentum \(\vec P\) of the two particle system 
(i.e. it holds for a moving or a stationary frame),
and for either degenerate or non-degenerate particles in each channel.
We follow closely the approach of Ref.~\cite{Kim:2005},
which presented a generalization of L\"uscher's single-channel
quantization condition to a moving frame.
Indeed, we find that the most general form of the final result,
given in Eq.~(\ref{eq:mdepres}), is identical in form to that of
Ref.~\cite{Kim:2005} (modulo some minor changes in notation).

After deriving the general quantization condition in Sec.~\ref{sec:multchan}
we restrict our considerations to the simplified situation
in which only s-wave scattering is included.
We focus on the case with two channels
(suggestively labeled $\pi\pi$ and $K\overline K$),
although we also provide the generalization to more than two channels.
In the infinite volume theory, the two-channel system is described by 
a \(2 \times 2\) \(S\)-matrix which, due
to unitarity and symmetry, is determined by three real parameters 
[see Eq.(\ref{eq:spar}) below]. 
We use a particular parametrization of
\(S\) to rewrite our quantization condition in a convenient, pure real
form [Eq.~(\ref{eq:repar})]. 
We explain how our result is equivalent to that of Ref.~\cite{Bernard:2010}
in the case of a stationary frame.
As is discussed in
Refs.~\cite{Lage:2009,Bernard:2010,Doering:2011}, three independent pieces
of information are needed to determine the three independent
\(S\)-matrix parameters at each center of mass energy, \(E^*\). 
References~\cite{Bernard:2010} and \cite{Doering:2011} discuss in
some detail the prospects for using either twisted boundary conditions
or uneven box sizes for this purpose. We restrict ourselves here to
an alternative approach, also mentioned in Refs.~\cite{Bernard:2010} and
\cite{Doering:2011}, of using
three different choices for the parameters $\{L,\vec P\}$, where
$L$ is the box size. 
(We assume a cubic box and periodic boundary conditions.)
The parameters $\{L,\vec P\}$ must be tuned such that there is a two-particle
state in the spectrum having the desired value of $E^*$.
In this way one obtains three independent conditions,
and can solve for the $S$-matrix parameters at the chosen value of $E^*$.

Turning now to the LL formula, we follow the same approach
as used by Lellouch and L\"uscher in Ref.~\cite{Lellouch:2000}.
Specifically, we add a $D$-meson to our two-channel system and
analyze the effect of an infinitesimal weak perturbation 
on the quantization condition. This yields a relation between
a finite-volume weak matrix element 
and a linear combination of the desired infinite-volume matrix elements. 
In Sec.~\ref{sec:LLgen} we present
a derivation of the relation which follows closely
the original LL work. In the final result, Eq.~(\ref{eq:llres}),
the coefficients relating finite and infinite volume matrix elements are
given in terms of the \(S\)-matrix parameters and their
derivatives, evaluated at the decay particle's mass. 
These can be calculated using the multiple-channel quantization
condition, as sketched above. 
It turns out that three different lattice matrix elements
are needed to separately determine the two infinite-volume matrix elements. 
Note that this is the same as the number 
needed to determine the $S$-matrix parameters. 
For more than two channels this correspondence no longer holds.

Since the $S$-matrix parameters and their derivatives are
ultimately determined from the spectral energies, it should
be possible to write a form for the generalized LL formula
in terms of the spectral energies and their derivatives alone.
We derive such a form in Sec.~\ref{sec:alt}.
The result, Eq.~(\ref{eq:finalform}), is probably more useful in practice
than Eq.~(\ref{eq:llres}).
The second derivation also brings out an important feature
of the generalized LL formula.
Finite-volume energy eigenstates in the coupled-channel theory 
can be written as linear combinations of
infinite volume $\pi\pi$ and $K\overline K$ states having 
(in our case) $\ell=0$ as well as the higher values of $\ell$ 
allowed by the cubic symmetry of the box.
The LL methodology is (as noted in the original paper)
simply a trick to determine the coefficients
of the relevant $\pi\pi$ and $K\overline K$ states.
This point has also been stressed recently by Ref.~\cite{Meyer:2012}
in a different context.
Our second derivation makes clear that, irrespective of
the details of the weak Hamiltonian, one always obtains
the same linear combination of $\pi\pi$ and $K\overline K$ states,
and that this feature holds for any number of channels.

The remainder of this article is organized as follows. In the
following section we give our derivation of the multiple channel
quantization condition. In Sec.~\ref{sec:pwave} we restrict to
\(s\)-wave scattering and derive a useful form of the condition. 
The multiple channel generalization of the LL formula is
then derived in Sec.~\ref{sec:LLgen}, and the
alternative derivation is presented in Sec.~\ref{sec:alt}. 
We conclude in Sec.~\ref{sec:conc}.
We include an appendix, in
which we discuss the generalization of Watson's theorem to two channels.

The generalization of L\"uscher's quantization formula 
to multiple channels for arbitrary $\vec P$ using field-theoretic
methods (the work described in our Secs.~\ref{sec:multchan} and
\ref{sec:pwave}) has also been considered by 
Briceno and Davoudi~\cite{RaulZohreh}.
Our results are in complete agreement (although we use a different
parametrization of the $S$-matrix).
Their paper is being released simultaneously with the present article.

\section{Multiple channel extension of quantization condition}
\label{sec:multchan}

In this section we derive an extension to multiple two-body channels of the 
L\"uscher quantization condition, which relates the infinite volume
scattering amplitudes to finite volume energy levels. We assume
throughout a cubic spatial volume with extent \(L\) and periodic
boundary conditions. The (Minkowski) time direction is taken to be infinite. 
The total momentum 
\begin{equation}
\vec P= \frac{2\pi \vec  n_P}{L} \hspace{30pt} ( \vec n_P \in \mathbb Z^3 )
\end{equation}
is fixed but arbitrary,
i.e. the quantization condition we derive holds for a ``moving frame''
as well as a stationary frame.
We first consider the case of only two open channels, describing
the extension to an arbitrary number of channels at
the end of this section.

We take each channel to contain two
massive, spinless particles. The particles of channel one are labeled
pions and are taken to be identical with mass $m_1=M_\pi$.
The particles of
channel two, called kaons, are taken non-identical, though still
degenerate, with mass $m_2=M_K$.
What we have in mind is that the first channel corresponds to the
$I=0$ $\pi \pi$ state, and the second to the 
$I=0$ $K \overline{K}$ state. 
Including both identical and non-identical pairs allows us
to display the factors of $1/2$ that appear in the former case.
We consider degenerate particles to simplify the presentation,
but describe the generalization to non-degenerate masses
at the end of this section.

For concreteness, and to match the physical ordering,
we take the pion to be lighter than the kaon. 
For our results to hold, we must assume that the thresholds
for three or more particles lie above the two kaon threshold.
If we assume a G-parity like symmetry, so that
only even numbers of pions can couple to a two-pion state,
then the ordering we need is
\begin{equation}
2 M_\pi < 2 M_K  < E^* < 4 M_\pi \,,
\label{eq:comconst}
\end{equation}
where \(E^*\) is the center of mass (CM) energy.
The only possible scattering events are then
\begin{equation}
\label{eq:decays}
\begin{split}
1 \rightarrow 1\!:& \hspace{10pt} \pi \ \pi \rightarrow \pi \ \pi 
\\ 
1 \rightarrow 2\!:& \hspace{10pt} \pi \ \pi \rightarrow K \ \overline K
\\ 
2 \rightarrow 1\!:& \hspace{10pt} K \ \overline K \rightarrow \pi\ \pi 
\\ 
2 \rightarrow 2\!:& \hspace{10pt} K \ \overline K \rightarrow K \ \overline K
\,.
\end{split}
\end{equation}
If $E^*$ drops below $2 M_K$, only the $\pi\pi$ channel is open
and the problem reduces to that discussed by 
L\"uscher~\cite{Luescher:1986n1, Luescher:1986n2, Luescher:1991n1,
Luescher:1991n2}.

The inequality $2 M_K< 4 M_\pi$ does not, of course, hold for
physical pions and kaons---the four and six pion thresholds occur
below that for two kaons. Nevertheless, the coupling to these
higher multiplicity channels is weak at low energies, and our
results should still hold approximately as long
as we are not too far above the two kaon threshold.
Indeed, it may be that, in the $I=0$ case,
the $\eta\eta$ channel becomes important before that with four or more
pions. If so, our formalism would still apply, generalized to
three channels as described below.
The approximation of ignoring channels with more than two particles
will become increasingly poor as the energy increases, and will
likely give only a rough guide by the $D$ mass. 
A qualitative indication of
this (ignoring differences in phase space)
is that the $f_0(1500)$ has a 50\% branching fraction to $4\pi$,
while the branches to 
$\pi\pi$, $K\overline{K}$ and $\eta\eta$ are
$\sim 35\%$, $9\%$ and $5\%$, respectively~\cite{rpp:2010}. 

The two channel quantization condition is 
obtained by a straightforward generalization of the 
single-channel approach of Ref.~\cite{Kim:2005}. 
To make this note somewhat independent of that reference, 
we reiterate some of the pertinent details. 
We begin by introducing a two body
interpolating field \(\sigma(x)\) (not necessarily local) which
couples to both channels. Following Ref.~\cite{Kim:2005} we then
define
\begin{equation}
\label{eq:fvc}
C_L(P) = \int_{L;x} e^{i(-\vec P \cdot \vec x + E x^{0})}\langle 0
\vert \sigma(x) \sigma^\dagger(0) \vert 0 \rangle
\end{equation}
where \(P = (E, \vec P)\) is the total four momentum of the two
particle system (in the frame where the finite volume condition is
applied), and
\begin{equation}
\int_{L;x} = \int_L d^4 x
\end{equation}
is the spacetime integral over finite volume. 
The relation to the CM energy used above is
\begin{equation}
E^* = \sqrt{E^2 - \vec P^{2}} \,.
\end{equation}
The poles of \(C_L\) give the energy spectrum of the
finite volume theory,
and thus the condition that \(C_{L}\) diverge
is precisely the quantization condition we are after.
\begin{figure}
\begin{center}
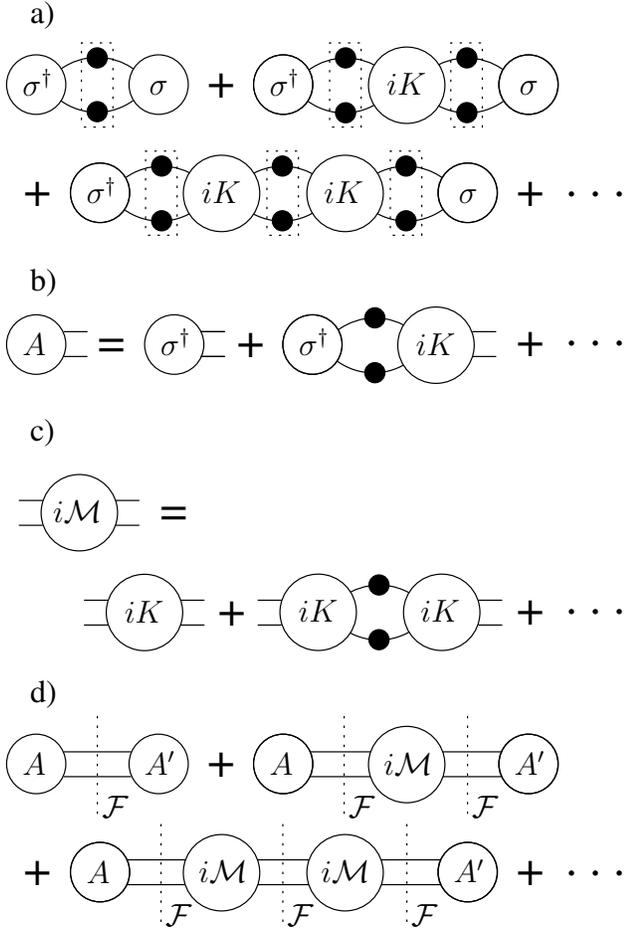
\caption{(a) The initial series of ladder diagrams which builds up
\(C_L\) [see Eq.~(\ref{eq:cl1})]. The Bethe-Salpeter kernels \(i K\)
are connected by fully dressed propagators. The dashed rectangle
indicates finite volume momentum sum/integrals. (b) and (c) The series
which build up the matrix element \(A\) and the scattering amplitude
\(i \mathcal M\). Note that these series contain only the momentum
integrals appropriate to {\em infinite} volume. (d) The resulting
series for the subtracted correlator [see Eqs.~(\ref{eq:fser}) and
(\ref{eq:mfseries})]. Each dashed vertical line indicates an
insertion of \(\mathcal F\), which carries the entire volume dependence
(neglecting exponentially suppressed dependence).}
\label{fig:fig1}
\end{center}
\end{figure}

To proceed to a more useful form of the condition, 
we follow Ref.~\cite{Kim:2005} and write $C_L$
in terms of the Bethe-Salpeter kernel, as illustrated
in Fig.~\ref{fig:fig1}(a):
\begin{multline}
\label{eq:cl1}
C_L(P) = \int_{L;q} \sigma_{j,q} \big [z^2 \Delta^2 \big]_{jk,q}
 \sigma^\dagger_{k,q} 
\\ 
+ \int_{L;q,q'} \sigma_{j,q} \big [z^2
 \Delta^2 \big]_{jk,q}i  K_{kl;q,q'} \big [z^2 \Delta^2 \big]_{lm,q'}
 \sigma^\dagger_{m,q'} + \cdots \,.
\end{multline}
The notation here is as follows.
Indices $j$, $k$, $l$ and $m$ refer to the channel, and
take the values $1$ or $2$.
The two particle intermediate states are summed/integrated
as is appropriate to finite volume
\begin{equation}
\int_{L,q}  = \frac{1}{L^3} \sum_{\vec q} \int \frac{d q^0}{2 \pi}\,. 
\end{equation}
The summand/integrand includes the product of two fully dressed
propagators
\begin{equation}
\big [z^2 \Delta^2 \big]_{ij,q}  = \delta_{ij} \eta_i
\big [z_i(q) \Delta_i(q) \big] \big[ z_i(P-q) \Delta_i(P-q) \big] \,,
\end{equation}
where 
\begin{align}
z_j(q) \Delta_j(q) &= 
\int d^4 x e^{i q x} \langle \phi_j(x) \phi_j(0) \rangle 
\\
\Delta_j(q) &= \frac{i}{q^2 - m_j^2 + i \epsilon} \,.
\end{align}
Here $\phi_1$ and $\phi_2$ are interpolating fields for
pions and kaons, respectively, chosen such that $z_j=1$ on shell.
\(\eta_1=1/2\) and \(\eta_2 = 1\) account for the symmetry factors
of the diagrams.
$K$ is related to the Bethe-Salpeter kernel
\begin{equation}
i K_{ij;q,q'} = i BS_{ij}(q, P-q,- q',- P+q')\,,
\end{equation}
with $BS_{ij}$ the sum of all amputated $j\to i$ scattering diagrams
which are two-particle-irreducible in the $s$-channel
(with particles of either type).
Finally, \(\sigma_{j,q}\) and $\sigma^\dagger_{j,q'}$
describe the coupling of the
operators $\sigma$ and $\sigma^\dagger$ to the two-particle
channel $j$. Their detailed form is not relevant; all we need to know
is that they are regular functions of \(q\).

We emphasize two important features of Eq.~(\ref{eq:cl1}).
First, it does not rely on any choice of interactions between the
pions and kaons, such as those predicted by chiral perturbation theory. 
All the quantities that enter can be written in terms of
non-perturbatively defined correlation functions.
Second, the kernel $i K$ and the propagator dressing function
$z$ have only exponentially suppressed dependence on the 
volume~\cite{Luescher:1986n2}.
Thus, if $L$ is large enough that such dependence
is negligible (as we assume hereafter), 
we can take $i K$ and $z$ to have their infinite-volume forms. 

The dominant power-law volume dependence enters 
through the momentum sums in the two-particle loops.
To extract this dependence, we
use the identity derived in Ref.~\cite{Kim:2005}, which
relates these sums for a moving frame to infinite-volume momentum
integrals plus a residue. Before stating the identity we recall the relevant
notation. 
For any four vector \(k^\mu = (k^0,\vec k)\) in the
moving frame, \(k^{\mu*} = (k^{0*},\vec k^*)\) is
the result of a boost to the CM frame. 
In particular, the total four-momentum $(E,\vec P)$ boosts to
$(E^*,\vec 0)$ in the CM frame.
We also need the quantities 
\begin{equation}
q_j^*=\sqrt{(E^*)^2/4-m_j^2} \,,
\end{equation}
which are the momenta of a pion ($j=1$) or
kaon ($j=2$) in the CM frame.
The identity then reads (no sum on \(i\) here):
\begin{multline}
\label{eq:sumid}
\int_{L;k} f(k) \eta_i \Delta_i(k) \Delta_i(P-k) g(k) = 
\\ 
\int_{\infty;k} f(k) \eta_i \Delta_i(k) \Delta_i(P-k) g(k)
\\ 
+ \int d \Omega_{q^*} d\Omega_{q^{*'}} \ 
f_i^*(\hat q^*) \ \mathcal F_{ii}(\hat q^*, \hat q^{'*})
g_i^*(\hat q^{'*})
\,,
\end{multline}
with
\begin{equation}
\int_{\infty;k}= \int \frac{d^4k}{(2\pi)^4} \,.
\end{equation}
We introduce two functions $f(k)$ and $g(k)$ to correspond to the momentum dependence
entering from the left and right of the loop integrals,
as well as that from the dressing functions [see Fig.~\ref{fig:fig1}(a)].
The functions $f$ and $g$ must have ultraviolet behavior that
renders the integral/sum convergent.
In addition, the branch cuts they contain, corresponding to four or more intermediate
particles, must be such that, after the $k^0$ contour
integration, they introduce no singularities for real $\vec k$.
This condition holds when $0< E^* < 4 M_\pi$.
The last line of (\ref{eq:sumid}) depends 
on the values of the functions $f$ and $g$ when the two particles
are on-shell, and thus only on the direction of the
CM momentum, $\hat q^*$. Specifically, if $q_i^\mu$ is the
moving frame momentum that boosts to the on-shell
momentum $(E^*/2, \vec q_i\,^*)$, then
\begin{equation}
\label{eq:fres}
f_i^*(\hat q^*) = f(q_i)\,, \quad
g_i^*(\hat q^*) = g(q_i)
\,.
\end{equation}
Finally the quantity ${\mathcal F}$, which depends on $q^*$, 
$L$ and the particle mass,
contains the power-law finite-volume dependence of the 
loop sum/integral.\footnote{%
The result (\ref{eq:sumid}) is equivalent to Eqs.~(41-42)
of Ref.~\cite{Kim:2005}, although we have done some further
manipulations to the last line of (\ref{eq:sumid})
to bring it into a matrix form.
Also, we have included a factor of $\eta_i$ in $\CF$, rather
than keeping it explicitly as in Ref.~\cite{Kim:2005}.}
Its form is given below in Eqs.~(\ref{eq:fco})-(\ref{eq:cdef}).
Note that it is diagonal in channel space, i.e. it cannot
change pions into kaons. It can, however, insert angular momentum,
due to the breaking of rotation symmetry by the cubic box.

The key point of the identity is that the difference between 
finite and infinite volume 
integrals depends on {\em on-shell} values of the integrand, 
allowing the finite-volume dependence to be expressed in terms
of physical quantities. 
Applying the identity to each loop integral in Fig.~\ref{fig:fig1}(a), 
one then rearranges the series by grouping terms with the
same number of insertions of ${\mathcal F}$.
The volume-independent term with no ${\mathcal F}$ insertions is
of no interest, since it does not lead to poles.
Thus we drop it and consider the difference
\begin{equation}
\label{eq:fser}
C_{\rm sub}(P) \equiv C_L(P) - C_{\infty}(P) \,.
\end{equation}
In the remaining diagrams with \(\mathcal F\) insertions, all terms to
the left of the first ${\mathcal F}$ and to the right of the last are
grouped and summed into new endcaps which we label \(A_j\) and
\(A'_j\) [see Fig.~\ref{fig:fig1}(b)]. These quantities equal certain matrix
elements of the interpolating field \(\sigma\)~\cite{Kim:2005}
\begin{align}
\label{eq:adef}
A_j(\hat k^*) & \equiv \langle \vec k^*,-\vec k^*;\ j;\ out \vert
\sigma^\dagger(0) \vert 0 \rangle_{\vert \vec k^* \vert = q^*_j} \\
\label{eq:apdef}
A'_j(\hat k^*) & \equiv  \langle 0 \vert \sigma (0)\vert \vec
k^*,-\vec k^*;\ j;\ in \rangle_{\vert \vec k^* \vert = q^*_j} \,.
\end{align}
In contrast to ~\cite{Kim:2005} we include no wavefunction
renormalization factors, because our single particle interpolating
fields satisfy on-shell renormalization conditions. Having summed up
the ends the next step is to do the same for the series which appears
between adjacent ${\mathcal F}$ insertions [Fig.~\ref{fig:fig1}(c)]. 
As indicated
in the figure, this series generates the infinite volume scattering
amplitude \(i \mathcal M_{ij}\). We thus deduce an alternative series
for \(C_{sub}\) built from \(A\), \(A'\) and \(i \mathcal M\)s, all
connected by \(\mathcal F\)s [Fig.~\ref{fig:fig1}(d)].

We stress that the analysis just performed is a straightforward
generalization of the single channel analysis of Ref.~\cite{Kim:2005}.
All that has changed is that
${\mathcal F}$ and $\CM$ are now $2\times2$
matrices in channel space, and $A$ and $A'$ vectors.

To proceed, we decompose $A$, $A'$, \(\mathcal M\) and
\(\mathcal F\) in spherical harmonics, defining coefficients via
\begin{align}
A_j(\hat k^*) & \equiv \sqrt{4 \pi} A_{j;\ell, m} Y_{\ell, m}(\hat k^*)
\\ 
A'_j(\hat k^*) & \equiv \sqrt{4 \pi} A'_{j;\ell, m} 
Y^*_{\ell,m}(\hat k^*) 
\\
\begin{split}
\mathcal M_{ij}(\hat k^*, \hat k^{'*}) & \equiv \\ & \hspace{-20pt} 4
\pi \mathcal M_{ij; \ell_1,m_1;\ell_2,m_2} Y_{\ell_1,m_1}(\hat k^*)
Y^*_{\ell_2,m_2}(\hat k^{'*})
\end{split} \\
\label{eq:fco}
\begin{split}
\mathcal F_{ij}(\hat k^*, \hat k^{'*}) & \equiv \\ & \hspace{-20pt} -
\frac{1}{4 \pi} F_{ij; \ell_1,m_1;\ell_2,m_2} Y_{\ell_1,m_1}(\hat k^*)
Y^*_{\ell_2,m_2}(\hat k^{'*}) \,,
\end{split}
\end{align}
where a sum over all \(\ell\)'s and \(m\)'s is implicit.
The factors of $4\pi$ are present so that we match the conventions
of Ref.~\cite{Kim:2005}. They imply, for example, that for a purely s-wave
amplitude, $\CM$ is the same in the two bases (for the $4\pi$
cancels with the two spherical harmonics).
The kinematical factor $F$ is given in Ref.~\cite{Kim:2005}
(aside from the above-noted factor of $\eta_i$)
and takes the form\footnote{%
An additional difference from Ref.~\cite{Kim:2005} is the 
appearance of $\Re q^*_i$ rather than $q^*$. This is discussed
in the next section.}
\begin{multline}
\label{eq:fdef}
F_{ij; \ell_1, m_1; \ell_2, m_2} 
\equiv \delta_{ij} F_{i; \ell_1, m_1; \ell_2, m_2} 
\\
= \delta_{ij} \eta_i \bigg [
\frac{\Re q^*_i}{8 \pi E^*} \delta_{\ell_1 \ell_2} \delta_{m_1 m_2} 
\\
\hspace{-10pt} - \frac{i}{2 E^*} \sum_{\ell,m} \frac{\sqrt{4
\pi}}{q_i^{*\,\ell}} c^P_{\ell m}(q_i^{*\,2}) \int d \Omega\;
Y^*_{\ell_1,m_1} Y^*_{\ell,m} Y_{\ell_2,m_2} \bigg] \,.
\end{multline}
Here the volume-dependence enters through the sums\footnote{%
We are slightly abusing the notation here for the sake of clarity.
$c^P_{\ell m}$ depends not only on $q^{*\, 2}$ but also on $m_i$,
but we keep the latter dependence implicit.
The dependence is made explicit at the end of this section.}
\begin{multline}
\label{eq:cdef}
c^P_{\ell m}(q^{*\, 2}) = \frac{1}{L^3} \sum_{\vec k}
\frac{\omega_k^*}{\omega_k} \frac{e^{\alpha(q^{*\, 2}-k^{*\,2})}}
{q^{*\, 2}-k^{*\, 2}} k^{*\, \ell} \sqrt{4 \pi} Y_{\ell, m}(\hat k^*)
\\ 
- \delta_{\ell 0} \ \ \mathcal P \! \int \frac{d^3  k^*}{(2\pi)^3} 
  \frac{e^{\alpha(q^{*\, 2}-k^{*\, 2})}}{q^{*\, 2}-k^{*\, 2}} 
\,,
\end{multline}
with $\omega_k=\sqrt{\vec k^2+m_i^2}$ being the energy of a particle
with momentum $\vec k$, and $\omega_k^*$ the energy after boosting to the
CM frame.
The properties of these sums are discussed in Ref.~\cite{Kim:2005}.

We are now in a position to write down the final result.
The series indicated in Fig.~\ref{fig:fig1}(d) gives
\begin{align}
\label{eq:mfseries}
C_{\rm sub}(P) &= - \sum_{n=0}^\infty
A' F \left[ - i \mathcal M F \right]^n A \,,
\\
&= - A' \frac1{F^{-1} + i \CM} A\,.
\label{eq:resum}
\end{align}
Here all indices are left implicit and may be restored in the
obvious way. For example,
\begin{multline}
A' F \mathcal M F A = A'_{i;\ell_1,m_1} F_{ij;\ell_1,m_1;\ell_2,m_2}
\\ \mathcal M_{jk;\ell_2,m_2;\ell_3,m_3} F_{kl;\ell_3,m_3;\ell_4,m_4}
A_{l,\ell_4,m_4} \,.
\end{multline}
As \(C_\infty\) has no poles in the region of \(E^*\) that we consider
(below \(4 M_\pi\)), the poles in \(C_L\) must match the poles in
\(C_{\rm sub}\). The desired quantization condition is then just
that the matrix between \(A'\) and \(A\) have a divergent
eigenvalue. This may be written as
\begin{equation}
\label{eq:mdepres}
\det \left(F^{-1} + i \mathcal M\right) = 0\,,
\end{equation}
where we recall that the matrices now act in the product of
the two-dimensional channel space and the infinite-dimensional
angular-momentum space. More precisely, $F$ is diagonal in
channel space but has off-diagonal elements between different
angular momentum sectors (as allowed by the symmetries of the
cubic box and the  momentum $\vec P$), while $\CM$ is diagonal
in angular momentum but off-diagonal in channel space.

Equation (\ref{eq:mdepres}) is the main result of this section. 
It has exactly the same form as that for the single channel
given in Ref.~\cite{Kim:2005} (aside from the change of notation in which
symmetry factors are contained in $F$ rather than kept explicit).
The generalization to more than two two-particle channels is now
immediate. As long as \(E^*\) is kept below the four particle
threshold of the lightest particle the arguments above go through in
the same manner. One need only extend the values of the
channel indices, taking care to include the appropriate symmetry
factor $\eta_j$ for each channel.
The final result then has exactly the form of Eq.~(\ref{eq:mdepres}).

To make the formal expression (\ref{eq:mdepres}) useful in practice
one assumes that there is some \(\ell_{\rm max}\), above which the
partial wave amplitudes are negligible
\begin{equation}
\mathcal M^{\ell > \ell_{\rm max}}_{ij} = 0 \,.
\end{equation}
One can then show that, although $F$ couples $\ell\le \ell_{\rm max}$
to $\ell > \ell_{\rm max}$, the projection contained in
$\CM$ is sufficient to collapse the required determinant
to that in the $\ell\le\ell_{\rm max}$ subspace.
The argument for this result is given for one channel in
Ref.~\cite{Kim:2005} and generalizes trivially to the
multiple channel case.
Thus one finds that Eq.~(\ref{eq:mdepres}) still  
holds, but with \(\mathcal M\) and \(F\) now
understood to be finite dimensional matrices both in channel space
and in the partial wave basis, with $\ell$ running up to \(\ell_{max}\).

To conclude this section we comment briefly on two generalizations of
the result. We first consider the case when not just a single
\(\sigma\) but rather a set of operators \(\{\sigma^a\}\) is of interest. 
This is likely to be the case in practice since multiple operators may
be needed to find combinations with good overlaps with the finite-volume
eigenstates.
If there are \(n\) such operators,
then \(C_L\) generalizes to an \(n \times n\) matrix:
\begin{equation}
C^{ab}_{L}(P) = \int_{L;x} e^{i(- \vec P \cdot \vec x + E x^0)} \langle
0 \vert \sigma^a(x) \sigma^{\dagger b}(0) \vert 0 \rangle \,.
\end{equation}
The generalization of Eq.~(\ref{eq:resum}) is effected by replacing
\(A'\) with an \(n \times 2\) matrix
\begin{equation}
\begin{pmatrix} A'_1 & A'_2 \end{pmatrix} 
\longrightarrow \begin{pmatrix} A^{'a=1}_1 & A^{'a=1}_2 
\\ A^{'a=2}_1 & A^{'a=2}_2 \\ \vdots & \vdots \end{pmatrix}
\end{equation}
and \(A\) with a \(2 \times n\) matrix
\begin{equation}
\begin{pmatrix} A_1 \\  A_2 \end{pmatrix} 
\longrightarrow 
\begin{pmatrix} A^{b=1}_1 & A^{b=2}_1 & \cdots 
\\ A^{b=1}_2 & A^{b=2}_2 & \cdots \end{pmatrix} \,.
\end{equation}
The key point, however, is that the matrix between $A'$ and $A$
is unchanged, so that the quantization condition (\ref{eq:mdepres})
is unaffected.
This is as expected, since the operators used to couple
to states cannot affect the eigenstates themselves.

The second generalization is to the 
case of non-degenerate particles. 
The expressions given above remain valid 
as long as one makes three changes.
First, the symmetry factors $\eta_i$ become unity for all non-degenerate
channels.
Second, $q_i^*$ in Eqs.~(\ref{eq:fdef}) is replaced by the solution of
\begin{equation}
E^* = \sqrt{q_i^{*\,2} + M_{ia}^2} + \sqrt{q_i^{*\,2} + M_{ib}^2}
\,,
\end{equation}
which is the CM momentum when the channel
contains particles of masses \(M_{ia}\) and \(M_{ib}\).
Third, when evaluating $c^P_{\ell m}$ using Eq.~(\ref{eq:cdef}),
one should use one of the masses $M_{ia}$ or $M_{ib}$ when
determining $\omega_k$, $\omega_k^*$ and $\vec k^*$.
One can show that both choices lead to the same result.

The third change emphasizes that the kinematic functions
$c^P_{\ell m}$ depend not only on $q_i^*$ but also on the
particle masses. This can be made explicit by rewriting them
in terms of a generalization of the zeta-function introduced
in Ref.~\cite{Kari:1995}. The result 
is~\cite{Davoudi:2011,Fu:2011,Leskovec:2012,RaulZohreh}
\begin{align}
c_{\ell m}^P(q^{*\,2}) &=
-\frac{\sqrt{4\pi}}{\gamma L^3}
\left(\frac{2\pi}{L}\right)^{\ell-2}
{\cal Z}_{\ell m}^P[1;(q^* L/2\pi)^2]
\label{eq:zetafunction}
\,,
\\
{\cal Z}_{\ell m}^P[s;x^2]
&=
\sum_{\vec n} \frac{r^\ell Y_{\ell m}(\hat r)}{(r^2-x^2)^s}
\,,
\end{align}
where $\gamma=E/E^*$,
$\vec n$ runs over integer vectors, and
$\vec r$ is obtained from $\vec n$ by
$r_\parallel=\gamma^{-1}[n_\parallel- c\vec n_P]$
and
$r_\perp=n_\perp$, where parallel and perpendicular are
relative to $\vec P$, and
$2 c =(1+(M_{1a}^2-M_{1b}^2)/E^{*\, 2})$.
The sum in ${\cal Z}_{\ell m}$ can be regulated by taking
$s > (3+\ell)/2$ and then analytically continuing to $s=1$.
This shows that mass dependence enters through 
the difference\footnote{%
The apparent lack of symmetry under the interchange
$M_{ia} \leftrightarrow M_{ib}$ can be understood as follows.
One can show that ${\cal Z}_{\ell m}^P \to (-)^\ell {\cal Z}_{\ell m}^P$
under this interchange (so that for degenerate masses the zeta-functions
for odd $\ell$ vanish~\cite{Kari:1995}). 
This sign flip for odd $\ell$ must hold also for the $c^P_{\ell m}$, 
and it does because the interchange of masses leads to 
$\vec k^* \to -\vec k^*$ at the pole. 
The sign flip is canceled in
the expression for ${\cal F}$, Eq.~(\ref{eq:fco}), since the product
$Y_{\ell_1,m_1}(\vec k^*) Y^*_{\ell_2,m_2}(\vec k^{'*})$
also changes sign. This is because, when $\ell$ is odd, the
integral over $d\Omega$ in the definition of $F$, Eq.~(\ref{eq:fdef}),
enforces that $\ell_1+\ell_2$ is odd. The overall effect is
that the quantization condition is symmetric under mass interchange,
as it must be.} 
$M_{ia}^2-M_{ib}^2$.
One can derive (\ref{eq:zetafunction}) by generalizing
the method used for the degenerate case in Ref.~\cite{Kim:2005}.

\section{Multiple-channel quantization condition for s-wave scattering}
\label{sec:pwave}

For the remainder of this article we focus on the simplest case,
\(\ell_{max}=0\), in which only s-wave scattering is significant.  In
this section we determine the explicit form for the finite-volume
quantization condition when there are two channels. We also present
compact forms for the condition when an arbitrary number of two
particle channels are open. 

With only s-wave scattering,
the two channel quantization condition takes the form
\begin{multline}
\label{eq:sonly}
[(F^s_1)^{-1} + i \mathcal M^s_{11}][(F^s_2)^{-1}+i \mathcal M^s_{22}]
\\ - [i \mathcal M^s_{12}][i \mathcal M^s_{21}] = 0 \,,
\end{multline}
where
\begin{align}
F^s_i & = \eta_i \left [\frac{\Re q^*_i}{8 \pi E^*} - \frac{i}{2E^*} 
c^{P}_i \right] 
\\ 
c^P_i & \equiv c^P(q_i^{*\,2}) \equiv c^P_{00}(q_i^{*\,2}) \,,
\end{align}
and the superscript on $F$ and $\CM$ is a reminder that only
$\ell=0$ contributes.

To simplify Eq.~(\ref{eq:sonly}), and in particular to re-express it
as an equation between real quantities, it is useful to recall
first the single-channel analysis.
This has the additional benefit of showing how the two-channel result
collapses to the known single-channel result in the appropriate kinematic
regime, namely
\begin{equation}
2 M_\pi < E^* < 2 M_K \,.
\end{equation}
In this regime $q^*_2$ becomes imaginary,
and the second channel contributes
negligibly because \(c^P\) [Eq.~(\ref{eq:cdef})]
becomes exponentially volume-suppressed and
\(\mathrm{Re\,} q^*\) in $F_2$ [Eq.~(\ref{eq:fdef})] vanishes.\footnote{%
The appearance of $\Re q^*$ rather than $q^*$
in $F_i$ can be understood by reviewing the derivation of \(F\) in
Ref.~\cite{Kim:2005}. The term enters as the
difference between principal part and \(i \epsilon\)
prescriptions. When \(q^*\) is imaginary there is no pole and
different ways of regulating give the same result.}
Sending \(F_2\to 0\) we find that the quantization condition becomes
\begin{equation}
\label{eq:sonech}
[\mathcal M_{11}^s]^{-1} = \eta_1 \left[ - \frac{i q_1^*}{8 \pi E^*} - \frac{1}{2 E^*} c^{P}(q_1^{*\,2}) \right] \,.
\end{equation}
Note that the pion momentum \(q^*_1\) is real for the energy region
considered.

Naively one might think that Eq.~(\ref{eq:sonech}) gives two
conditions, the separate vanishing of the real and imaginary
parts. This is not the case, however,
because the vanishing of the imaginary part
is a volume-independent condition which is guaranteed to hold by the
unitarity of the \(S\)-matrix. This can be seen by
expressing $\CM$ in terms of the real phase shift \(\delta(q^*)\),
\begin{equation}
\label{eq:pshift}
\mathcal M_{11}^s = \frac{8 \pi E^*}{\eta_1 q_1^*} \left[
\frac{e^{2i\delta(q_1^*)}-1}{2i}\right] = \left[ \frac{\eta_1q_1^*}{8 \pi E^*}
\left[\cot \delta(q_1^*) - i \right] \right]^{-1} \,.
\end{equation}
Here \(e^{2i\delta}\) is the one dimensional unitary \(S\)-matrix in
the partial wave basis. Given Eq.~(\ref{eq:pshift}), it is manifest
that the imaginary part of Eq.~(\ref{eq:sonech}) holds automatically.
The real part of (\ref{eq:sonech}) then gives the moving frame
generalization of the L\"uscher result in the familiar partial wave
form~\cite{Luescher:1991n1,Kim:2005,Christ:2005}
\begin{equation}
\label{eq:pshiftcond}
\tan \delta(q_1^*) = - \tan \phi^P(q_1^*) \,,
\end{equation}
where
\begin{equation}
\tan \phi^P(q^*) = \frac{q^*}{4 \pi} \left [c^{P}(q^{*\,2}) \right]^{-1} \,.
\end{equation}

We now return to the CM energies for which both channels
are open, $2 M_K < E^* < 4 M_\pi$,
and generalize Eq.~(\ref{eq:pshiftcond}).
The first step is to recall the relationship between
the scattering amplitude and the \(S\)-matrix.
Unitarity implies that
\begin{equation}
\label{eq:Munitarity}
\CM^s-\CM^{s\,\dagger} = i \CM^{s\,\dagger} P^2 \CM^s
\,,
\end{equation}
where $P^2$ is a diagonal matrix containing the
phase-space factors, whose square root is
\begin{equation}
\label{eq:pdef}
 P = \frac{1}{\sqrt{4 \pi E^*}}
\begin{pmatrix}
\sqrt{\eta_1 q_1^*} & 0 \\ 0 & \sqrt{\eta_2 q_2^*} \\
\end{pmatrix} 
\,.
\end{equation}
We note that, when expressed in terms of $q^*$, the form of
$P$ is still valid if the two particles in the channel are non-degenerate. 
We also note that the form (\ref{eq:Munitarity}) holds for an
arbitrary number of two-particle, s-wave channels, with $P$
generalized in the obvious way.

The solution to the unitarity relation is
\begin{equation}
\label{eq:twochm}
i \mathcal M^s = P^{-1} \left( S^s - 1 \right) P^{-1}
\end{equation}
where $S^s$ is a dimensionless, unitary $2\times 2$ matrix.
To proceed, we need to parametrize $S^s$.
First we note that $S^s$ can be taken to be symmetric.
This is because of the T-invariance of the strong interactions,
together with the fact that angular momentum eigenstates have
definite T-parity (in our case, positive).
Thus in the $2\times 2$ case, $S^s$ is determined by three real parameters.
We use the ``eigenphase convention'' of Blatt and
Biedenharn~\cite{Blatt:1952},
\begin{equation}
\label{eq:spar}
S^s =
\begin{pmatrix}
 \ccc_\epsilon & - \! \sss_\epsilon \\ \sss_\epsilon & \ccc_\epsilon
\end{pmatrix}
\begin{pmatrix}
e^{2 i \delta_{\alpha}} & 0 \\ 0 & e^{2 i \delta_{\beta}}
\end{pmatrix}
\begin{pmatrix}
\ccc_\epsilon & \sss_\epsilon \\ - \! \sss_\epsilon & \ccc_\epsilon
\end{pmatrix} \,,
\end{equation}
where the notation \(\sss_x = \sin x\) and \(\ccc_x = \cos x\) will be used throughout. The three real parameters \(\delta_\alpha\),
\(\delta_\beta\), and \(\epsilon\) generalize
the single \(\delta\) which appears in the one channel case. The
parameter \(\epsilon\) quantifies the mixing between the mass
eigenstates of channels one and two (the pions and kaons) and the
\(S\)-matrix eigenstates. The phases \(\delta_\alpha\) and
\(\delta_\beta\) are, for arbitrary \(\epsilon\), associated with both
channels. Of course, in the limit \(\epsilon \rightarrow 0\) they
reduce, respectively, to the phase shifts of pion and kaon elastic
scattering.

Substituting the form of \(S^s\) into Eq.~(\ref{eq:twochm}) and then
placing this in Eq.~(\ref{eq:sonly}) and simplifying, we deduce\footnote{%
We emphasize that the physical content of Eq.~(\ref{eq:repar}),
namely that there is a relation between scattering amplitudes
and energy levels, does not depend on the 
parametrization chosen for the matrix $S^s$.
This is clear either from Eq.~(\ref{eq:sonly}) or
from Eq.~(\protect\ref{eq:Sform}) below.
An advantage of our choice of parametrization 
is that it shows that Eq.~(\ref{eq:sonly})
only implies one real condition (rather than two), an observation
which must hold for any parameterization.
We also note that the freedom to independently change the phases of $\pi\pi$
and $K\overline{K}$ states, which leads to $S^s\to U^\dagger S^s U$,
with $U$ a {\em diagonal} unitary matrix, does not change
the quantization condition, as can be seen most easily
from Eq.~(\protect\ref{eq:Sform}) below.
}
\begin{multline}
\label{eq:repar}
\left [ \tan \delta_\alpha + \tan \phi^P(q^*_1) \right ] \left [ \tan
\delta_\beta + \tan \phi^P(q^*_2) \right ] \\ + \sin^2 \! \epsilon \,
\left [ \tan \delta_\alpha - \tan \delta_\beta \right ] \left [\tan
\phi^P(q^*_1) - \tan \phi^P(q^*_2) \right ] \\ = 0 \,.
\end{multline}
This is the main result of this section. 
One can use it in one of two ways: to predict the spectrum given
knowledge of the scattering amplitude from experiment or a model,
or to determine the $S$-matrix parameters from a lattice calculation
of the spectrum.
In the former case, we note that all quantities appearing
in (\ref{eq:repar}), i.e.
\(\delta_\alpha\), \(\delta_\beta\), \(\epsilon\), $q^*_i$
and $\phi^P$, are functions of $E^*$.
One can thus search, at 
given spatial extent \(L\) and total momentum $\vec P$, 
for values of \(E^*\) which satisfy Eq.~(\ref{eq:repar}). 
If the condition holds for a particular
\(E^*_k\), then
\begin{equation}
\label{eq:Ekdef}
E_k(L;\vec n_P) = \sqrt{E^{*2}_k + \vec P^{\,2}}
\end{equation}
is in the spectrum of the finite-volume moving-frame
Hamiltonian. Here we choose to write $E_k$ as a function
of $\vec n_P$ rather than $\vec P$, since, in practice,
it is the former quantity which is held fixed as one varies $L$.

The second use of (\ref{eq:repar}) is the most relevant for
the discussion in subsequent sections.
For a given choice of $E^*$, one finds, through a lattice
calculation, three pairs $\{L,\vec n_P\}$ for which there
is a spectral line $E_k$ such that $E^*_k$ 
[defined in Eq.~(\ref{eq:Ekdef})]
is equal to $E^*$.
One could use a fixed $\vec n_P$ and consider multiple spectral lines
(the simplest choice conceptually), 
or use three different choices of $\vec n_P$
(probably more practical since one would not need to determine so
many excited levels).
In either case, one ends up with three versions of Eq.~(\ref{eq:repar}),
all containing the desired quantities
\(\delta_\alpha(E^*)\), \(\delta_\beta(E^*)\)
and \(\epsilon(E^*)\), but having different values of the $\phi^P(q^*_j)$.
Solving these equations one determines, rather
than the angles themselves, the quantities \(\tan \delta_\alpha\),
\(\tan\delta_\beta\), and \(\sin^2 \epsilon\) at CM energy $E^*$. 
For our discussion we therefore restrict
\begin{equation}
\label{eq:range}
\delta_{\alpha,\beta} \in [-\pi/2,\pi/2] \mathrm{\ \ and\ \ } \epsilon
\in [0,\pi/2] \,.
\end{equation}
Having determined the restricted phases over a range of energies, one
can afterward relax the constraints in order to build continuous
curves as a function of energy. We direct the reader to
Refs.~\cite{Bernard:2010,Doering:2011} for discussion of other methods
for extracting the three scattering parameters.

We emphasize that Eq.~(\ref{eq:repar}) has a very intuitive form. If
\(\delta_{\alpha} = \delta_{\beta}\) or \(m_1 = m_2\) or \(\epsilon =
0\) then the second line vanishes and the result reduces to two
copies of the one channel quantization condition
[Eq.~(\ref{eq:pshiftcond})]. To see that this makes sense, note that
for identical phase shifts, the \(\epsilon\) matrix commutes through
the phase matrix and we recover two uncoupled channels. Similarly if
the masses are degenerate then the eigenstates of the \(S\)-matrix
will also be mass eigenstates leading to the decoupled form. Finally,
the decoupling for \(\epsilon=0\) is an obvious property of the
parametrization [Eq.~(\ref{eq:spar})].

An alternative solution to the unitarity relation (\ref{eq:Munitarity})
can be given in terms of the $K$-matrix used in Ref.~\cite{Bernard:2010}.
Specifically, (\ref{eq:Munitarity}) is satisfied if
\begin{equation}
\left({\cal M}^{s}\right)^{-1} = M - i P^2/2
\,,
\end{equation}
with $M$ any real symmetric $2\times 2$ matrix.
If we set
\begin{equation}
M = \frac1{8\pi E^*} \|\sqrt\eta\|\, K^{-1}\, \|\sqrt\eta\|
\,,
\end{equation}
[where double bars denote a diagonal matrix, so that
$\|\eta\|={\rm diag}(\sqrt{\eta_1},\sqrt{\eta_2})$],
and further set $\vec P=0$,
then one can show that the two-channel quantization condition
given above can be manipulated into the form given in
Eq.~(3.5) of Ref.~\cite{Bernard:2010} in terms of the
real, symmetric matrix $K$.

We now generalize Eq.~(\ref{eq:repar}) to $N$ s-wave channels.
As noted above, the form of the unitarity relation (\ref{eq:Munitarity})
holds for any $N$, and the same is true for
its solution (\ref{eq:twochm}).
In the latter, the $N$ channel \(S\)-matrix can be parametrized as\footnote{%
The remainder of this paper is limited to the \(s\)-wave, so we
drop the superscript \({}^s\) hereafter.}
\begin{equation}
\label{eq:gensmat}
S = R^{-1} \, \big \| e^{2i\delta} \big \| \, R \,,
\end{equation}
with \(R\) an \(SO(N)\) matrix,
and
\begin{equation}
\label{eq:diagdelt}
\big \| e^{2i\delta} \big \| = {\rm diag} \big [e^{2i\delta_\alpha},
e^{2i \delta_\beta}, \cdots \big ] \,.
\end{equation}
Together with
Eqs.~(\ref{eq:gensmat}) and (\ref{eq:diagdelt}) one needs the 
\(N \times N\) generalization of \(F\), which has
been discussed in the previous section.
From these definitions one can straightforwardly work out the 
quantization condition for \(N\) coupled channels.

We conclude this section by describing two additional ways of writing
the quantization condition, both of which make the higher channel
generalization especially clear. Observe first that, for any number of
open channels,
\begin{equation}
F^{-1} = P^{-1} \, \big \| 1-e^{-2i\phi} \big \| \, P^{-1}
\,.
\end{equation}
Combining this with (\ref{eq:twochm}), it follows that
\begin{equation}
\label{eq:XYrel}
F^{-1} + i \CM = P^{-1} \Big [ S - \big \| e^{-2i\phi} \big \| \Big ]
P^{-1} \,.
\end{equation}
Since $P^{-1}$ has no singularities in the
kinematic regime we consider, the
quantization condition can be rewritten as
\begin{equation}
\det \Big [ S - \big \| e^{-2i\phi} \big \| \Big ] = 0 \,.
\label{eq:Sform}
\end{equation}
This form shows that the symmetry factors cancel from the
quantization condition in general.
Although Eq.~(\ref{eq:Sform}) looks like it will lead to one complex
and thus two real conditions, it turns out that it leads only to a
single real condition. This follows from the identity
\begin{multline}
\big \| 1 + i t_\phi \big \|  \times \Big [ S -
\big \| e^{-2i\phi} \big \| \Big ] \times \Big [ R^{-1} \big \| 1 - i
t_\delta \big \| R \Big ]
\\ 
= 2 i \Big[ R^{-1} \big \| t_\delta \big \| R + \big\| t_\phi \big \|\Big] \,,
\end{multline}
where \(t_x = \tan x\). It gives a manifestly real rewriting of the
quantization condition,
\begin{equation}
\det \Big [ R^{-1} \big \| t_\delta \big \| R + \big \| t_\phi \big \|
\Big ] = 0 \,.
\end{equation}
This form leads directly to the result (\ref{eq:repar}) 
in the two channel case,
and collapses to the single-channel result (\ref{eq:pshiftcond}) 
for any channel that decouples from the rest.
If any of the channels contain non-degenerate particles,
this enters only through the values of the kinematic
functions $t_\phi$, as discussed in the previous section.

\section{Multiple channel extension of the Lellouch-L\"uscher formula}
\label{sec:LLgen}

Having found the two channel quantization condition, we are now in a
position to work out the two channel generalization of the LL
formula which relates weak matrix elements in finite and
infinite volume. The derivation follows the original work by Lellouch
and L\"uscher, Ref.~\cite{Lellouch:2000}, which was extended to a
moving frame by Refs.~\cite{Kim:2005,Christ:2005}.

We begin by introducing a third channel which is decoupled from the
original two. This contains a single particle,
which we call a \(D\)-meson, whose mass satisfies
\begin{equation}
M_D>2 M_\pi, 2 M_K\,.
\end{equation}
We next introduce a weak perturbation to the Hamiltonian density
\begin{equation}
\mathcal H(x) \rightarrow \mathcal H(x) + \lambda \mathcal H_W(x) \,,
\end{equation}
where \(\lambda\) is a parameter which can be varied freely and can,
in particular, be taken arbitrarily small. The perturbation \(\mathcal H_W\) 
is defined to couple channels one and two (pions and kaons) to
the third (\(D\)-meson) and nothing more.
It is convenient to choose it to
be invariant under time reversal (T) symmetry. 
The generalization to perturbations which are not 
T invariant is described  at the end of the section. 

Consider now the finite volume spectrum, first in the absence of the
perturbation. The spectrum of two-particle states with $\vec P=2\pi \vec n_P/L$
is determined by Eq.~(\ref{eq:repar}).
It is \(L\)-dependent and \(L\) can therefore be tuned to make one of the
levels, call it \(k_D\), degenerate with the energy of a single (moving) \(D\)
meson
\begin{equation}
E_{k_D} = \sqrt{M_D^2 + \vec P^{\,2}}
\end{equation}
With $L$ fixed in this way, we now turn on the weak interaction. At
leading order in degenerate perturbation theory this changes the
energies to
\begin{equation}
E^{(1)} = E^{(0)} \pm \lambda V \vert M_W \vert
\end{equation}
where $V=L^3$, $E^{(0)} = E_{k_D}$,
and the finite-volume matrix element is
\begin{align}
 M_W & = {}_L\langle k_D \vert \mathcal H_W(0) \vert D \rangle_L.
\end{align}
The subscripts $L$ on the states indicate that they are normalized
to unity, unlike the relativistically normalized infinite volume states.
Superscripts \({}^{(1)}\) are used throughout this section to indicate
that the quantity includes both the leading order and the 
\(\mathcal O(\lambda)\) correction, while superscripts \({}^{(0)}\) 
indicate the unperturbed quantity. The effect of the
perturbation may also be written in terms of the CM energy as
\begin{align}
\label{eq:energyshift}
E^{*(1)} & = M_D \pm \lambda \Delta E^* \\ \Delta E^* & =
\frac{E^{(0)} V \vert M_W \vert}{M_D} \,.
\end{align}

\begin{figure}
\begin{center}
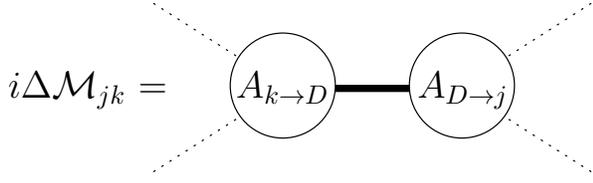
\caption{The diagram giving rise to 
the amplitude perturbation \(\Delta \mathcal M\) [See Eq.~(\ref{eq:DeltaM})].
\label{fig:LL}}
\end{center}
\end{figure}

Of course, in addition to affecting the finite volume energy spectrum,
the weak perturbation also changes the infinite volume scattering
amplitudes. The leading order effect is generated by the 
diagram of Fig.~~\ref{fig:LL}, which gives
\begin{equation}
\mathcal M^{(1)} = \mathcal M^{(0)} \mp \lambda \Delta \mathcal M
\end{equation}
with
\begin{equation}
\label{eq:DeltaM}
\Delta \mathcal M_{j,k} = 
\frac{\langle j \vert \mathcal H_W(0)
\vert D \rangle \langle D \vert \mathcal H_W(0) \vert k \rangle}
{2 E^{(0)} V \vert M_W \vert} \,.
\end{equation}
The indices \(j\) and \(k\) run over the two
two-particle channels. 
This perturbation may equivalently be
represented through shifts in \(\delta_\alpha\), \(\delta_\beta\)
and \(\epsilon\)
\begin{align}
\delta^{(1)}_\alpha(E^*) & = \delta^{(0)}_\alpha(E^*) \pm \lambda
\Delta \delta_\alpha(E^*) \\ \delta^{(1)}_\beta(E^*) & =
\delta^{(0)}_\beta(E^*) \pm \lambda \Delta \delta_\beta(E^*) \\
\epsilon^{(1)}(E^*) & = \epsilon^{(0)}(E^*) \pm \lambda \Delta
\epsilon(E^*) \,,
\end{align}
The explicit forms of the perturbed phases are given in
Eqs.~(\ref{eq:deldelal})-(\ref{eq:deleps}) below.

The calculation now proceeds as follows.
When the quantities
\begin{equation} 
{\delta^{(0)}_\alpha(E^*)},\ {\delta^{(0)}_\beta(E^*)},\ \mathrm{and}\
{\epsilon^{(0)}(E^*)}
\end{equation}
are placed in the quantization condition [Eq.~(\ref{eq:repar})],
the condition is satisfied by construction at \(E^{*(0)}=M_D\). 
Alternatively if one places
\begin{equation} 
\delta^{(1)}_\alpha(E^*),\ \delta^{(1)}_\beta(E^*),\ \mathrm{and}\
\epsilon^{(1)}(E^*)
\end{equation}
into the same condition then it must be satisfied when
evaluated at the perturbed CM energy \(E^{*(1)}\), but only to linear order in \(\lambda\). 
The constant order term in the
\(\lambda\) expansion is just the unperturbed condition, and so it is
the vanishing of the \(\mathcal O (\lambda)\) term that is of
interest. 
The condition that this term vanish gives the relation
between finite and infinite volume weak matrix elements that we are
after.

The only detail left to discuss, before substituting into the
quantization condition and expanding in \(\lambda\), is the explicit
forms of the amplitude corrections to \(\delta_\alpha\),
\(\delta_\beta\) and \(\epsilon\). Before these are found it is useful
to determine the constraints on the infinite volume matrix
elements which arise from Watson's theorem. As shown in
App.~\ref{app:weakform}, time reversal invariance and unitarity
constrain the matrix elements to be such that the
following two quantities are real:
\begin{align}
\label{eq:v1def}
\hspace{-5pt}v_1 & = e^{-i \delta_\alpha} \left[\sqrt{q^*_{1}\eta_1} A_{D
\rightarrow \pi \pi} \ccc_\epsilon + 
\sqrt{q^*_{2}\eta_2} A_{D \rightarrow K K} \sss_\epsilon \right] 
\,, \\
\label{eq:v2def}
\hspace{-5pt} v_2 & = e^{-i \delta_\beta} \left[ - \sqrt{q^*_{1}\eta_1} 
A_{D \rightarrow \pi \pi} \sss_\epsilon + \sqrt{q^*_{2}\eta_2} 
A_{D \rightarrow K K} \ccc_\epsilon \right]\,.
\end{align}
Here 
\begin{equation}
A_{D \rightarrow \pi \pi}  \equiv 
\langle \pi \pi \vert \mathcal H_W(0) \vert D \rangle \,,
\end{equation}
and similarly for the $K \overline K$ case, normalized so that the decay
rates to each channel are
\begin{equation}
\Gamma_{D\to j} = \frac{q^*_j \eta_j}{8 \pi M_D^2} |A_{D\to j}|^2
= \frac{1}{2 M_D} P_{jj}^2 |A_{D\to j}|^2\,.
\label{eq:Gamma}
\end{equation}
This relation holds also if the particles in a channel 
are non-degenerate (requiring $\eta=1$).
All energy-dependent parameters in (\ref{eq:v1def}) and (\ref{eq:v2def}), 
i.e. \(\delta_\alpha\), \(\delta_\beta\), \(\epsilon\) and \(q^*_j\),
are to be evaluated at \(E^*=M_D\). 

The results (\ref{eq:v1def}) and (\ref{eq:v2def}) hold when
the phases of the states are chosen so that the $S$-matrix is
symmetric (as is possible given T invariance).
This does not determine the signs of the two matrix elements,
and these signs are unphysical. More precisely, the relative sign
ambiguity is the same as the ambiguity in the sign of $\epsilon$,
so once we have fixed the latter to be positive, the relative sign
is physical. The overall sign remains unphysical, and can be chosen,
for example to set \(v_1 \geq 0\),

Inverting the relations (\ref{eq:v1def}) and (\ref{eq:v2def}) yields
\begin{align}
\label{eq:pieltxt}
A_{D \rightarrow \pi \pi}  
& = \frac1{\sqrt{q^{*}_{1}\eta_1}} \left[ 
v_1 e^{i \delta_\alpha} \ccc_\epsilon - v_2 e^{i \delta_\beta}
\sss_\epsilon\right]
\\
\label{eq:keltxt}
A_{D \rightarrow K K} 
& = \frac1{\sqrt{q^{*}_{2}\eta_2}} \left[ 
v_1 e^{i\delta_\alpha} \sss_\epsilon + v_2 e^{i \delta_\beta} \ccc_\epsilon
\right ]
\,.
\end{align}
Inserting these in $\Delta \CM$, Eq.~(\ref{eq:DeltaM}),
and using the relation between $\CM$ and $S$,
Eq.~(\ref{eq:twochm}), and the parametrization of $S$, 
Eq.~(\ref{eq:spar}), we find that perturbations to $\delta_\alpha$,
$\delta_\beta$ and $\epsilon$ are real. This is a consistency
check on the calculation (or an alternative derivation of the
Watson's theorem constraint). Specifically, we find
\begin{align}
\label{eq:deldelal}
\Delta \delta_{\alpha} & = - \mathcal N v_1^2 \\
\label{eq:deldelbet}
\Delta \delta_{\beta} & = - \mathcal N v_2^2 \\
\label{eq:deleps}
\Delta \epsilon & = - \mathcal N \frac{v_1 v_2}{\ccc_\alpha
\ccc_\beta(\ttt_\alpha-\ttt_\beta)}
\end{align}
where \(\ttt_\alpha = \tan \delta_\alpha\), etc., and
\begin{equation}
\mathcal N = \frac{1}{16 \pi E^{(0)} M_D V \vert M_W\vert} \,.
\end{equation}

We now have all the ingredients to substitute into the quantization
condition and determine the LL generalization. We
emphasize that, when the expansion in $\lambda$
is performed, \(\delta_\alpha\),
\(\delta_\beta\) and \(\epsilon\) each contribute both from
the amplitude corrections of
Eqs.~(\ref{eq:deldelal})-(\ref{eq:deleps}) and from the shift in
the energy, (\ref{eq:energyshift}).
The other contributions arise from the energy
dependence of \(\phi_i = \phi^P(q^*_i)\). Substituting and
simplifying, we find the main result of this section
\begin{equation}
\label{eq:llres}
\mathcal C_{1} v_1^2 + \mathcal C_{2} v_2^2 + \mathcal C_{1 2} v_1 v_2
= \mathcal C_{M^2} \vert M_W \vert^2
\end{equation}
where
\begin{align}
\label{eq:c1def}
\begin{split}
\mathcal C_{1} & = \frac{\pi}{16} \frac{\ttt_1 + \ttt_2 + 2
\ttt_\beta+ (\ttt_2 - \ttt_1) (1 - 2 \sss_\epsilon^2)}{\ccc_\alpha^2}
\end{split}\\
\begin{split}
\mathcal C_{2} & = \frac{\pi}{16} \frac{\ttt_1 + \ttt_2 + 2
\ttt_\alpha + (\ttt_1 - \ttt_2) (1 - 2 \sss_\epsilon^2)}{\ccc_\beta^2}
\end{split}\\
\label{eq:c12def}
\begin{split}
\mathcal C_{12} & = \frac{\pi}{4} (\ttt_1 - \ttt_2)
\frac{\sss_{\epsilon}\! \ccc_{\epsilon}}{\ccc_\alpha \! \ccc_\beta}
\end{split} \\
\label{eq:cmdef}
\begin{split}
\mathcal C_{M^2} & = \frac{\pi^2 M_D V^2 (E^{(0)})^2}{2} 
\bigg [\frac{\ttt'_1}{q_1^*}
\left(\ttt_2 + \ttt_\beta +  (\ttt_\alpha -
\ttt_\beta)\sss_\epsilon^2 \right) \\ & \hspace{50pt} + \frac{\ttt'_2}{q_2^*}
\left(\ttt_1 + \ttt_\alpha +  (\ttt_\beta -
\ttt_\alpha) \sss_\epsilon^2 \right) \\ & \hspace{50pt}+ \frac{4 \ttt_\alpha'}{M_D}
\left(\ttt_2 + \ttt_\beta + (\ttt_1 - \ttt_2) \sss_\epsilon^2 \right)
\\ & \hspace{50pt}+ \frac{4 \ttt_\beta'}{M_D} \left (\ttt_1 +
\ttt_\alpha + (\ttt_2 - \ttt_1) \sss_\epsilon^2 \right) \\ &
\hspace{50pt}+ \frac{4 \,{\sss^{2}_\epsilon}'}{M_D} (\ttt_1 - \ttt_2)
(\ttt_\alpha - \ttt_\beta) \bigg] \,,
\end{split}
\end{align}
and where we use the additional notation
\begin{equation}
\ttt_i = \tan \phi^P[q^*_i] \,.
\end{equation}
All quantities are evaluated at the \(D\) mass,
and we have dropped the superscript ${}^{(0)}$.
The primes on \(\phi_i\) indicate derivatives with respect to
\(q^*_i\) while those on \(\delta_\alpha\), \(\delta_\beta\) and
\(\epsilon\) indicate derivatives with respect to \(E^*\). 
In each case, these are the natural variables on which the
quantities depend.
We have checked that this formula reduces to (two copies of)
the single-channel LL result if $\epsilon\to 0$.

We now describe how the result (\ref{eq:llres}) can be used
in practice. A lattice calculation yields the finite-volume
matrix element $|M_W|$, and the aim is to determine the
infinite-volume matrix elements $A_{D\to\pi\pi}$ and
$A_{D\to KK}$. Using the generalized quantization condition
(\ref{eq:repar}) for three different spectral lines
(all chosen to have $E^*=M_D$)
one can determine \(\delta_\alpha\), \(\delta_\beta\) and
\(\epsilon\) as described in the previous section.
Repeating the procedure at
a slightly different energy allows a numerical determination
of the required derivatives.
One now evaluates $|M_W|$ at the degenerate point on one of the
spectral lines. The knowledge of the $S$-matrix parameters and
their derivatives, together with the value of \(L\),
allows one to calculate the values of the four \(\mathcal C\)'s
[Eqs.~(\ref{eq:c1def})-(\ref{eq:cmdef})]. 
Combined with the value of \(|M_W|\), one then finds from 
Eq.~(\ref{eq:llres}) a quadratic constraint on $v_1$ and $v_2$.
Repeating the procedure for a second spectral line
gives an independent constraint, which allows for
the determination of $v_1$ and $v_2$ up to a two-fold ambiguity
corresponding to the unknown relative sign. 
Finally, repeating for a third spectral line resolves the sign ambiguity.
With $v_1$ and $v_2$ determined in this way,
one can obtain the infinite-volume matrix elements using
Eqs.~(\ref{eq:pieltxt}) and (\ref{eq:keltxt}).
Although this procedure is rather elaborate, we note that 
(for the case of two channels) three
spectral lines are needed both for the determination
of the parameters of the $S$-matrix and of the LL factors.

We conclude this section by commenting that Eq.~(\ref{eq:llres})
factors as
\begin{equation}
\label{eq:cvfactors}
{\rm sgn}(\mathcal C_1)(c_1 v_1 + c_2 v_2)^2 = 
\mathcal C_{M^2} \vert M_W \vert^2
\end{equation}
where
\begin{equation}
\label{eq:lilcdef}
c_1 = \sqrt{\vert \mathcal C_1 \vert} \hspace{25pt} 
c_2 = {\rm sgn}[\mathcal C_1 \mathcal C_{12}] \sqrt{\vert \mathcal C_2 \vert} \,.
\end{equation}
The only new information encoded in Eqs.~(\ref{eq:cvfactors}) and
(\ref{eq:lilcdef}) relative to Eq.~(\ref{eq:llres}) is that
\begin{equation}
4 \mathcal C_1 \mathcal C_2 = \mathcal C_{12}^2 \,,
\end{equation}
which can be shown to hold by applying Eq.~(\ref{eq:repar}) to Eqs.~(\ref{eq:c1def})-(\ref{eq:c12def}).
Although the factorized form (\ref{eq:cvfactors}) is simpler,
it does not reduce the number of values of $L$ that are needed because 
there remains a sign ambiguity (from the square root) at each $L$.
What it does make clear, however, is
that the generalized LL condition will fail when the signs of
\(\mathcal C_1\) and \(\mathcal C_{M^2}\) are opposite.
Presumably this cannot happen for physical values of the phase shifts.
We stress that this issue also arises in the original one-channel set-up,
where the LL formula only makes sense if
\begin{equation}
\frac{d (\delta+ \phi^P)}{d q^*} > 0\,.
\end{equation}
We return to these sign constraints in the next section.

The form (\ref{eq:cvfactors}) also allows one to write the LL condition 
as a factored form in terms of the desired matrix elements,
\begin{equation}
\label{eq:melfact}
\vert c_\pi A_{D \rightarrow \pi \pi} + c_K A_{D \rightarrow K K}
\vert^2 = \vert \mathcal C_{M^2} \vert \vert M_W \vert ^2 \,,
\end{equation}
where $c_\pi$ and $c_K$ are complex, and can be determined from the
above results.
As this equality holds for any T-invariant form of weak perturbation and for any
decay particle, it must imply a relation between finite
and infinite volume states
\begin{equation}
\label{eq:cpicKdef}
{}_L\langle k_D \vert \propto c_\pi \langle \pi \pi,{\rm out} \vert 
+ c_K \langle K \overline K, {\rm out} \vert + \dots\,.
\end{equation}
Here the ellipsis indicates the $\pi\pi$ and $K \overline K$ states
of higher angular momentum which are needed to satisfy the 
periodic boundary conditions.
Indeed, as noted in the original derivation of Ref.~\cite{Lellouch:2000},
the use of the $D$-meson is simply a trick to obtain the
normalization factors.\footnote{%
In the one-channel case, an alternative line of argument has
been developed for obtaining the LL relation,
based on matching the density of two-particle states in
finite and infinite volumes~\cite{Lin:2001}. 
In the present case, we do not see how to use this approach
to determine the relative normalization, $c_K/c_\pi$, 
of the two components in the finite volume state. 
Thus we think that this approach could provide
only a consistency check.}
It follows that Eq.~(\ref{eq:melfact}) 
must also hold for perturbations which are not T-invariant.

The appearance of the linear combination in Eq.~(\ref{eq:cpicKdef})
can be better understood from an alternative derivation of the LL
formula, to which we now turn.

\section{Alternative derivation of Lellouch-L\"uscher formula}
\label{sec:alt}

In this section we present a different derivation of the two channel
LL relation which has the following advantages: (a) it does not
require determining the shifts $\Delta\delta_\alpha$, $\Delta
\delta_\beta$ and $\Delta\epsilon$, but rather works directly with the
change in $\CM$; (b) it gives one directly a condition with the
factored form, proportional to the left hand side of
Eq.~(\ref{eq:melfact}); (c) it allows one to rewrite the LL condition
in a simpler form in which the only inputs required are the
derivatives of the energies with respect to $L$ along the
three spectral lines. This form is likely to be more practical.

We work directly with the condition $\det(F^{-1} + i \CM)=0$, and keep
results for general number of channels, $N$,
 as far as possible. We begin by defining
\begin{eqnarray}
\CX &=& F^{-1} + i \CM \\ \CY &=& S - \big \| e^{-2i\phi} \big \| \,.
\end{eqnarray}
and recall from Eq.~(\ref{eq:XYrel}) that
\begin{equation}
\CX = P^{-1} \CY P^{-1} \,.
\end{equation}
The quantization condition $\det\CX=0$ is equivalent to $\CX$ having
an eigenvector with vanishing eigenvalue.  We label this eigenvector
$\overrightarrow e^X$.
Note also that the symmetry of $\CX$ implies $(\overrightarrow
e^{X})^{Tr}=\overleftarrow e^X$ is a left eigenvector, also with zero
eigenvalue.

Now we can formulate the LL condition in a relatively
compact form. As above, let $\CM^{(0)}$ be the scattering amplitude at
CM energy $E^*=M_D$. Similarly, let $F^{(0)}$ be the finite-volume
factor at this CM energy and for one of the values of box size $L$ for
which the quantization condition holds.  Then for
\begin{equation}
\CX^{(0)} \equiv(F^{(0)})^{-1}+ i \CM^{(0)}\,,
\end{equation}
we have
\begin{equation}
\overleftarrow e^{X} \CX^{(0)} \overrightarrow e^{X} = 0 \,.
\end{equation}
Now, while holding $L$ fixed, we change the energy by $\pm
\lambda\Delta E = \pm \lambda V |M_W|$ and change $\CM$ to
$\CM^{(0)}\mp \lambda \Delta \CM$, and require that the quantization
condition still hold.  Thus we have, to linear order in $\lambda$,
\begin{equation}
\det(\CX^{(0)} +\lambda \Delta \CX) = 0 \,,
\end{equation}
where
\begin{equation}
\Delta \CX = \pm \Delta E \frac{\partial \CX}{\partial E}\Bigg|_L \mp
i \Delta\CM \,.
\end{equation}
It follows that there must be a new eigenvector of the form
\begin{equation}
\overrightarrow e^{X} + \lambda \Delta \overrightarrow e^{X}
\end{equation}
which is annihilated by the perturbed matrix. From the \(\mathcal
O(\lambda)\) term in
\begin{equation}
\left [ \overleftarrow e^{X} + \lambda \Delta \overleftarrow e^{X}
\right ] \left[ \CX^{(0)} +\lambda \Delta \CX \right ] \left [
\overrightarrow e^{X} + \lambda \Delta \overrightarrow e^{X} \right ]
=0 \,,
\end{equation}
we deduce
\begin{equation}
 \overleftarrow e^X \Delta\CX \overrightarrow e^X = 0 \,.
\end{equation}
Using the explicit form of $\Delta\CX$ this becomes
\begin{equation}
\Delta E\; \overleftarrow e^X \frac{\partial \CX}{\partial E}\Bigg|_L
\overrightarrow e^X = \overleftarrow e^X i \Delta\CM\; \overrightarrow
e^X \,,
\end{equation}
where the derivative is evaluated at $E^*=M_D$.

We can slightly simplify this result by expressing the left hand side
in terms of $\CY$ rather than $\CX$, and thus removing factors of
$P^{-1}$. The point is that, when the derivative acts on the $P^{-1}$
factors in $\CX$, the contribution to the left hand side vanishes,
since one can still act (either to the left or the right) on the
zero-eigenvector. Thus we can rewrite the condition in terms of the
zero eigenvector of $\CY$, which is
\begin{equation}
\overrightarrow e^{Y} = P^{-1} \overrightarrow e^X\,.
\label{eq:eYdef}
\end{equation}
The new form is
\begin{equation}
\Delta E\; \overleftarrow e^Y \frac{\partial \CY}{\partial E}\Bigg|_L
\overrightarrow e^Y = 
\overleftarrow e^X [ i \Delta\CM ] \overrightarrow e^X \,.
\label{eq:Ycond}
\end{equation}

We now focus on the $2\times2$ case.  To proceed, we need the explicit
form for $\overrightarrow e^Y$, which is given, up to an overall
normalization factor, by
\begin{equation}
\overrightarrow e^Y = \begin{bmatrix} 1 \\ z e^{i(\phi_2-\phi_1)} \end{bmatrix}
\end{equation}
where $z$ is the real quantity
\begin{equation}
z= t_\epsilon \frac{\sin(\delta_\beta+\phi_1)}{\sin(\delta_\beta+\phi_2)} \,.
\end{equation}
It is clear from the form of Eq.~(\ref{eq:Ycond}) and
the relation (\ref{eq:eYdef}) that the
normalization of $\overrightarrow e^Y$ is irrelevant and so we have
chosen a relatively simple unnormalized form.

We evaluate the right hand side of Eq.~(\ref{eq:Ycond})
using the form of $\Delta \CM$, Eq.~(\ref{eq:DeltaM}). 
It is immediately apparent that
the result factorizes, given that $\Delta \CM$ is an outer
product. This will hold for all $N$. In the $N=2$ case we have
\begin{equation}
\overleftarrow e^X [ i \Delta\CM ] \overrightarrow e^X 
= i e^{-2i\phi_1}\frac{M_\infty^2}{8 \pi M_D E^{(0)} V |M_W|} \,,
\label{eq:DeltaMres}
\end{equation}
where
\begin{multline}
 M_\infty =\\
 e^{i\phi_1} \sqrt{q_1^*\eta_1}\, A_{D \rightarrow \pi \pi} + z\, e^{i\phi_2} \sqrt{q_2^*\eta_2}\, A_{D \rightarrow K K}
\,.
\label{eq:Minfty_res}
\end{multline}
Here we have used the assumed T-invariance of $\CH_W$.
We have pulled out the phase $e^{-2i\phi_1}$ so that $M_\infty$ is real. Its
reality is not obvious, but can be established using the results
derived from Watson's theorem and given in 
App.~\ref{app:weakform}.
In particular, an algebraic exercise shows that
\begin{multline}
M_\infty = \\
\sin(\phi_1-\phi_2) \left [ -v_1
\frac{c_\epsilon}{\sin(\delta_\alpha\!+\!\phi_2)} +v_2
\frac{s_\epsilon}{\sin(\delta_\beta\!+\!\phi_2)}\right ] \,,
\end{multline}
and we recall that the quantities $v_1$ and $v_2$ [defined above in
Eqs.~(\ref{eq:v1def})-(\ref{eq:v2def})] are real.

The result (\ref{eq:Minfty_res}) makes clear that, for any choice of
$\CH_W$, one ends up with the matrix element to a given (complex)
linear combination of $\langle\pi\pi|$ and $\langle K \overline K|$ states, since
all the factors ($\phi_1$, $\phi_2$ and $z$) are determined by
$E^{(0)}$ and $L$. Indeed, what the LL method has allowed us to do is
determine the coefficients of the s-wave $\langle\pi\pi|$ and
$\langle K \overline K|$ components within the finite-volume state. 
As mentioned above, this
decomposition has nothing to do with $\CH_W$, and thus we can use the
result for any $\CH_W$, including one involving T-violation.
By comparing the result (\ref{eq:Minfty_res})
to the general decomposition of the finite-volume state,
Eq.~(\ref{eq:cpicKdef}), we can read off the ratio of the 
coefficients,
\begin{equation}
\frac{c_K}{c_\pi} = e^{i(\phi_2-\phi_1)}\, z\, 
\sqrt{\frac{q_2^*\eta_2}{q_1^*\eta_1}}
\,.
\end{equation}
It is interesting that the relative phase between $c_K$ and $c_\pi$
is determined by the kinematic phases $\phi_j$.
Given the form of $\Delta\CM$, and the fact that, in Eq.~(\ref{eq:DeltaMres}),
it is sandwiched between $\overleftarrow e^X$ and $\overrightarrow e^X$,
it follows that the zero eigenvector itself gives the relative
size of the $\pi\pi$ and $K\overline{K}$ contributions:
\begin{equation}
\overrightarrow e^X \propto \begin{pmatrix} c_\pi \\ c_K \end{pmatrix}\,.
\end{equation}
This illustrates in a direct way that the linear combination which
appears is completely independent of the form of ${\mathcal H}_W$,
since the eigenvector of $\CX$ knows nothing about this perturbation.

Having discussed the right hand side of Eq.~(\ref{eq:Ycond}) 
in some detail we now turn to the left. 
Specifically, we show that it is possible to write the left-hand
side in terms of the derivative of the spectral energy with respect to $L$.
To motivate this form, we first recall that the LL result of the
previous section depends on \(\delta_\alpha\), \(\delta_\beta\) and
\(\epsilon\) and their derivatives, evaluated at \(E^* = M_D\).
As described in Sec.~\ref{sec:LLgen},
the three \(S\)-matrix parameters may be determined, using
Eq.~(\ref{eq:repar}),
by finding three different pairs \(\{L,\vec n_P\}\) for
which there is a spectral line $E_k(L;\vec n_P)$ satisfying
\(E_k^*(L;\vec n_P)=M_D\) [see Eq.~(\ref{eq:Ekdef})].
Furthermore, by slightly changing the three \(L\) values, one
can determine \(\delta_\alpha\), \(\delta_\beta\) and \(\epsilon\) at
slightly different energies and thus deduce the derivatives at
\(M_D\).

The point of reiterating these steps is to note that, since the
lattice simulation actually gives the energy spectrum as a function of
\(L\), it would be preferable if the LL result could be rewritten
directly in terms of the properties of the spectrum. In this way the
extra step of separately working out the phase shifts and their
derivatives would be avoided. This turns out to be possible, as we now
show.

We use the quantization condition in the form $\det\CY=0$.  To stay on
a spectral line $E_k(L;\vec n_P)$ as we vary $E$ away 
from the moving frame \(D\)-meson energy $E^{(0)}$, we need to vary $L$
in such a way that this condition remains fulfilled.  We note that,
while $F$ depends on both $E^*$ and $L$, $S$ depends only on
$E^*$. Thus we use $E^*$ and $L$ as independent variables.  Then the
condition to stay on a spectral line becomes
\begin{equation}
0= \overleftarrow e^Y \left[\Delta E^* \frac{\partial \CY}{\partial
E^*}\Bigg|_L + \Delta L \frac{\partial\CY}{\partial
L}\Bigg|_{E^*}\right] \overrightarrow e^Y \,,
\end{equation}
which leads to
\begin{equation}
\frac{d E_k^*}{d L}\Bigg|_{\rm line} = -
\frac{\overleftarrow e^Y \frac{\partial\CY}{\partial L}
\overrightarrow e^Y} {\overleftarrow e^Y \frac{\partial\CY}{\partial
E^*} \overrightarrow e^Y} \,.
\label{eq:dEk_dL}
\end{equation}
Here, in the left-hand side, the subscript ``line'' indicates that
the derivative is along a spectral line with fixed $\vec n_P$.

The key features of Eq.~(\ref{eq:dEk_dL}) are
that the denominator on the right-hand side is, up to a simple overall
factor, equal to the quantity appearing on the left hand side of the
Eq.~(\ref{eq:Ycond}), while the numerator is a kinematic factor.
Specifically, using
\begin{equation}
\overleftarrow e^Y \frac{\partial\CY}{\partial E^*} 
\overrightarrow e^Y = \frac{E^*}{E} \overleftarrow e^Y
\frac{\partial\CY}{\partial E} 
\overrightarrow e^Y \,,
\end{equation}
and
\begin{equation}
\frac{d E_k}{d L}\Bigg|_{\rm line} =
\frac{E_k^*}{E_k}\frac{d E_k^*}{d L}\Bigg|_{\rm line} -
\frac{\vec P^2}{EL}
\end{equation}
(which follows since $E^2=(E^*)^2 + (\vec P L)^2/L^2$ and because
$\vec P L$ is fixed along the spectral line), we find
\begin{equation}
\overleftarrow e^Y \frac{\partial\CY}{\partial E} 
\overrightarrow e^Y = 
- 2 i e^{-2i\phi_1} 
\frac{\frac{\partial \phi_1}{\partial L} 
      + z^2  \frac{\partial \phi_2 }{\partial L}}
{\frac{d E_k}{d L}\big|_{\rm line} + \frac{\vec P^2}{E L}} \,.
\end{equation}
Combining this with (\ref{eq:Ycond}) and (\ref{eq:DeltaMres}) we
conclude
\begin{eqnarray}
\frac{M_\infty^2}{16 \pi M_D E^{(0)} V^2 |M_W|^2} &=& -
\frac{\frac{\partial \phi_1 }{\partial L} + z^2 \frac{\partial \phi_2
}{\partial L} } {\frac{d E_k}{d L}\big|_{\rm line} +
\frac{\vec P^2}{E^{(0)} L}} \,.
\label{eq:finalform}
\end{eqnarray}
We thus have found an alternative form of the LL relation which is simpler than
Eq.~(\ref{eq:llres}), and also likely to be more practical.

The single channel version of (\ref{eq:finalform}) is instructive.
It can be written, using Eq.~(\ref{eq:Gamma}), in terms of the
decay rate:
\begin{equation}
\Gamma_{D\to \pi\pi} =
\frac{2 E^{(0)} V^2 |M_W|^2}{M_D} \left [ \frac{- \frac{\partial \phi }{\partial L} } 
{\frac{d E_k}{d L}\big|_{\rm line} + \frac{\vec P^2}{E^{(0)} L}} \right ]\,.
\end{equation}
This form holds both for identical and non-identical particles,
with the symmetry factor being contained in $\Gamma$.
It also sheds light on the sign constraints discussed in the
previous section. The right-hand side must be positive.
Based on numerical studies, we find that $\partial \phi/\partial L$
is always positive, implying that the denominator, which
is proportional to $d E_k^*/dL$, must be negative.

The same holds for the two-channel result, Eq.~(\ref{eq:finalform}).
In order for the right-hand side to be positive, the denominator must
be negative.
Since we could do the LL analysis on almost any spectral line,
this appears to imply that $d E_k^*/dL < 0$ in general.
The only exception is for a state with $E_k^*$ below the two particle
threshold. Such a state occurs, for example, as the lowest
energy state for $\vec P=0$ if there is an attractive interaction.
For such a state one has $d E_k^*/dL=dE_k/dL > 0$, i.e. of the ``wrong''
sign. But in this case the LL analysis does not apply, because the
particle lies below threshold in infinite volume.

\section{Conclusions}
\label{sec:conc}

We have presented two new results in this article.
First, a field-theoretic derivation of the generalization of L\"uscher's
quantization formula to the case of multiple strongly coupled
two-particle channels (where the particles are spinless).
Second, the generalization to multiple channels
of the Lellouch-L\"uscher formula relating finite-volume and
infinite-volume matrix elements.
We also have explained in some detail how, in the case of two channels,
one can use these results to determine the infinite-volume
decay amplitudes of a particle which is coupled by a weak
interaction to the two-body channels. 

As already noted in the introduction, this is but a step on
the way toward our ``dream'' application, namely the calculation of
$D^0\to\pi\pi$ and $D^0\to K\overline K$ amplitudes.
To achieve that goal, one will also need to include the channels
with four or more pions. These are significant once one approaches
the energy $M_D$. Work in this direction is underway.

An example where our formalism should be useful with minimal approximation
is the determination of the isospin breaking in $K\to\pi\pi$ decays.
Given the mass splitting between charged and neutral pions, there are
really two two-body channels to consider, and in this case the coupling
to the four pion channel is very small and can reasonably be neglected.

\section*{Acknowledgments}

This work is supported in part by the US DOE grant
DE-FG02-96ER40956. We thank Raul Briceno, Zohreh Davoudi,
Harvey Meyer and Ulf Meissner for discussions and comments on the manuscript.
This work was facilitated in part by the workshop 
``New Physics from Heavy Quarks in Hadron Colliders,''
which was  sponsored by the University of Washington and supported
by the DOE.

\appendix

\section{Two-channel Watson's theorem \label{app:weakform}}

In this appendix we work out the consequences of Watson's theorem
for the phases of the matrix elements of interest,
$\langle \pi \pi \vert \mathcal H_W(0) \vert D \rangle$
and
$\langle K \overline K \vert \mathcal H_W(0) \vert D \rangle$.
We assume at first that $\mathcal H_W$ is T invariant, and describe
the generalization to non-invariant Hamiltonians at the end.
We closely follow the textbook presentation given in 
Ref.~\cite{Weinberg}.

We consider the $3\times 3$ $S$-matrix with the three
states being the hypothetical $D$ meson (at rest) and the 
s-wave $\pi\pi$ and $K\overline{K}$ states. We assume that we are in the
kinematic regime described in the main text, so that the $3\times 3$
$S$-matrix is unitary. 
Although we introduce a weak coupling between
the $D$ and the two particle states, so that the $D$ is a resonance,
its width is of second-order in the weak interaction and thus can
be ignored at the linear order to which we work. Thus it is valid
to treat it as an asymptotic state.

Watson's theorem follows by breaking the $S$-matrix
into a strong part $S^{(0)}$ and a weak part $S^{W}$.
The strong part is T invariant, and, since we use states
which have definite (positive) T-parity, can be taken to be symmetric.
This fixes the phases of the $\pi\pi$ and $K\overline K$ states,
though not their overall signs.
Extending the dimensionless, strong-coupling \(S\)-matrix 
of Eq.~(\ref{eq:twochm}) to include the \(D\) gives 
\begin{equation}
S^{(0)} =
\begin{pmatrix}
1 &  0 \\ 0 & S^s 
\end{pmatrix}  \,,
\end{equation}
where $1$ is the $1\times 1$ identity and
$S^s$ is the $2\times 2$ s-wave $S$-matrix
given in (\ref{eq:twochm}). 
The weak part only contains couplings between the $D$ and
the two-particle states, and in $3\times3$ notation is
\begin{equation}
S^{W} =
\begin{pmatrix}
0 & S_{D,\pi \pi}^{W} & S_{D,K K}^{W} \\ S_{\pi \pi,D}^{W} &
0 & 0 \\ S_{K K,D}^{W} & 0 & 0
\end{pmatrix}\,.
\end{equation}
The assumed T invariance implies that it, too, is symmetric.
The non-zero elements of $S^{W}$
are proportional to the desired matrix elements
\begin{align}
\label{eq:DeltaSvsHW}
S_{j,D}^{W}
&= c P_{jj} \langle j |[-i \mathcal H_W(0)]| D\rangle
\,,
\end{align}
where $j=1,2$ runs over the $\pi\pi$ and $K\overline K$ channels,
$P$ is the square root of the phase space factor
defined in Eq.~(\ref{eq:pdef}),
and $c$ is a known real constant whose value will not be needed.

Unitarity of the complete $S$-matrix implies that the
terms linear in the weak interaction satisfy
\begin{equation}
i S^{W} = S^{(0)} \big(i S^{W} \big)^\dagger S^{(0)} \,.
\end{equation}
This implies that
\begin{equation}
i S_{j,D}^{W} = S^s_{jk} \big(i S_{D,k}^{W}\big)^* 
= S^s_{jk} \big(i S_{k,D}^{W}\big)^* 
\,,
\label{eq:keyWatson}
\end{equation}
where in the last step we have used the symmetry of $S^{W}$.
Using the explicit form for the two-channel $S$-matrix\footnote{%
For simplicity of presentation, we are here using $\delta_1=\delta_\alpha$
and $\delta_2=\delta_\beta$.}
\begin{equation}
S^s = R^{-1} 
\begin{pmatrix}
e^{2i \delta_{1}} & 0 \\ 0 & e^{2i\delta_{2}}
\end{pmatrix}
    R
\,,
\end{equation}
with
\begin{equation}
R =
\begin{pmatrix}
\ccc_\epsilon & \sss_\epsilon \\ - \sss_\epsilon & \ccc_\epsilon
\end{pmatrix} \,,
\end{equation}
we find
\begin{equation}
i R_{jk} S_{k,D}^{W} = 
e^{2i \delta_{j}}  \big(iR_{jk} S_{k,D}^{W}\big)^* 
\,.
\end{equation}
It follows that the phase of
$i R_{jk} S_{k,D}^{W}$ is $e^{i\delta_j}$.
This is the desired generalization of Watson's theorem to two channels.
Thus the quantities
\begin{equation}
v_j = e^{-i\delta_j} \frac1c \sqrt{4\pi E^*}\, i R_{jk} S_{k,D}^{W}
\end{equation}
are real. Using (\ref{eq:DeltaSvsHW}) we can rewrite the $v_j$
as in Eqs.~(\ref{eq:v1def}) and (\ref{eq:v2def}).

If the weak interaction is not T invariant, then
$S_{j,D}^{W}$ will contain some number of T-violating phases.
Since we are working to linear order in the weak interaction,
we can break up $\mathcal H_W$ into parts each with a single
T-violating phase and treat each separately. 
Each such part has an overall phase $e^{i\phi_T}$,
and the symmetry of the S-matrix is replaced by
\begin{equation}
S_{D,k}^{W}(\phi_T) = S_{k,D}^{W}(-\phi_T)
\,.
\end{equation}
However, if we first pull out the overall phase by hand, then
the symmetry of $\Delta S$ is restored, and Watson's theorem
applies to the residue.

\bibliographystyle{apsrev} 
\bibliography{ref} 

\end{document}